\documentclass{PoS}

\title{Fermi-point scenario for emergent gravity}

\ShortTitle{Fermi-point scenario for emergent gravity}

\author{\speaker{G.E. Volovik}
\\
       Low Temperature Laboratory, 
Helsinki University of Technology,
P.O.Box 2200, FIN-02015 HUT, Finland \\
L.D. Landau Institute for Theoretical Physics, 119334
 Moscow, Russia\\
        E-mail: \email{volovik@boojum.hut.fi}}

\abstract{Let us assume that gravity is an emergent low-energy phenomenon arising from a topologically stable defect in momentum space --  the Fermi point. What are the consequences?
We discuss the natural values of fermion masses and cosmological constant; flatness of the Universe; bounds on Lorentz violation; etc. }

\FullConference{From Quantum to Emergent Gravity: Theory and Phenomenology\\
		June 11-15 2007\\
		Trieste, Italy}

\begin{document}

\section{Natural values of physical quantities}

In emergent  physics the natural value  of a physical quantity means that this value naturally emerges in the effective low energy theory without fine tuning. Both in particle physics and condensed matter the natural value of a quantity depends on whether this quantity is determined by macroscopic or  microscopic physics, see Table \ref{NaturalValues}:

\begin{equation}
\matrix{
\begin{array}{lccc}
{\rm  physical~quantity} &~~{\rm natural~value} ~~ &~~{\rm dimensional~analysis}  ~~&~~{\rm observation}   \cr
\cr
\hline
\hline
\cr
 {\rm Newton ~constant}  &E_{\rm P}^{-2}   &E_{\rm P}^{-2}   &E_{\rm P}^{-2}    \cr
  {\rm running ~coupling ~constant}  & 1   &1   &\sim 1   \cr
    {\rm mass~of~Higgs~boson}          &E_{\rm P}   & E_{\rm P}        &\approx 0     \cr
    \cr
    \hline
    \cr
    {\rm temperature~of~Universe}  &0      &E_{\rm P}   &\approx 0  \cr
{\rm cosmological ~constant~\& ~vacuum~pressure}  &0   &E_{\rm P}^4 &\approx 0  \cr
 {\rm volume~of~Universe}          &\infty      &E_{\rm P}^{-3}       &\infty      \cr
\cr
 \hline
 \cr
      {\rm mass~of~elementary~particle}          &E_{\rm P} ~~{\rm or}~~0     & E_{\rm P}        &\approx 0     \cr
\cr
 \hline
 \cr
 \end{array}
 }
 \label{NaturalValues}
 \end{equation}
\vskip2mm

The first column in the Table \ref{NaturalValues} 
contains the natural values of the physical quantities. In the second column the estimates of these quantities are shown, which follow from dimensional analysis assuming that the role of the fundamental energy scale is played by Planck energy  $E_{\rm P} $. In the third column the observational values are given; here we neglect the magnitudes which are much smaller than the Planck scale values. For example, the observed masses of elementary particles, the upper limit for the mass of Higgs  boson,  the observed value of the cosmological constant, and the highest temperature in the Universe are many orders of magnitude  smaller than their Planck scale values, and thus are considered as almost identical zero.

Most of the quantities are determined by microscopic physics and are expressed in terms of the corresponding microscopic scale, which is the Planck scale $E_{\rm P}$ in our Universe or atomic scale in condensed matter systems.  An example  is the Newton constant $G=a_G E_{\rm P}^{-2}$. For emergent gravity, the dimensionless prefactor $a_G$  depends on the vacuum content and is of order unity in units $\hbar=c=1$ (in the Fermi-point scenario which we discuss here, $\hbar$ is the fundamental constant, while the parameter $c$ -- the maximum attainable speed of the low-energy particles -- is determined by the microscopic physics). In principle, the parameter  $a_G$ can be zero, but this requires fine-tuning between different scalar, vector and spinor fields in the vacuum. That is why the natural value of $G$ is $E_{\rm P}^{-2}$.
The natural value of the  mass of the Higgs boson is the Planck energy, 
 $M_{\rm Higgs}=a_H E_{\rm P}$. However, from the  condensed matter systems we know that the prefactor $a_H$ depends much on the complexity of the system, and can be exponentially reduced. This is what  happens to the transition temperature $T_c$ in superconductors and Fermi superfluids, which is exponentially suppressed almost in all systems except for the high-$T_c$ cuprates.
The running coupling constants $\alpha_n$ also falls into this category, since they depend on the ultraviolet cut-off together with the infra-red cut-off  $E_{\rm IR}$: $\alpha_n^{-1}\sim \ln (E_{\rm P}/E_{\rm IR})$ .  

Temperature,  pressure,  and volume belong to the category determined by macroscopic physics -- thermodynamics. These thermodynamic quantities do not depend on the micro-physics or on momentum-space  topology; they only depend on the environment. In the absence of forces from the environment, the pressure and temperature of any system relax to zero. The same should hold for the temperature of the Universe and for the vacuum pressure. The vacuum pressure is, with a minus sign, the cosmological constant, $\Lambda=\epsilon_{\rm vac}=-p_{\rm vac}$. Whatever is the vacuum content, and independently of the history of phase transitions in the quantum vacuum, the cosmological constant must relax to zero or to the small value which compensates the other partial contributions to the total pressure of the system: it is the total pressure of the system that must be zero in equilibrium.

Masses of elementary particles fall into a special category. The naive estimation tells us that these masses should be on the order of the Planck energy scale:  $M_{\rm expected} \sim E_{\rm P}\sim 10^{19}$ GeV. This highly contradicts observations: the observed masses of known particles are many orders of magnitude smaller, being below the electroweak energy scale $M_{\rm real}<E_{\rm ew}\sim 1$  TeV.  This  represents the main hierarchy problem. In the ``natural''   Universe, where all masses are of order $E_{\rm P}$, all fermionic degrees of freedom are completely frozen out because of the  Bolzmann factor $e^{-M/T}$, which is about $e^{-E_{\rm P} /E_{\rm ew}} \sim e^{-10^{16}}$  already at the temperature corresponding to the  highest energy reached in accelerators. There is no  fermionic matter in such a Universe.  

\begin{figure}
\centerline{\includegraphics[width=1.0\linewidth]{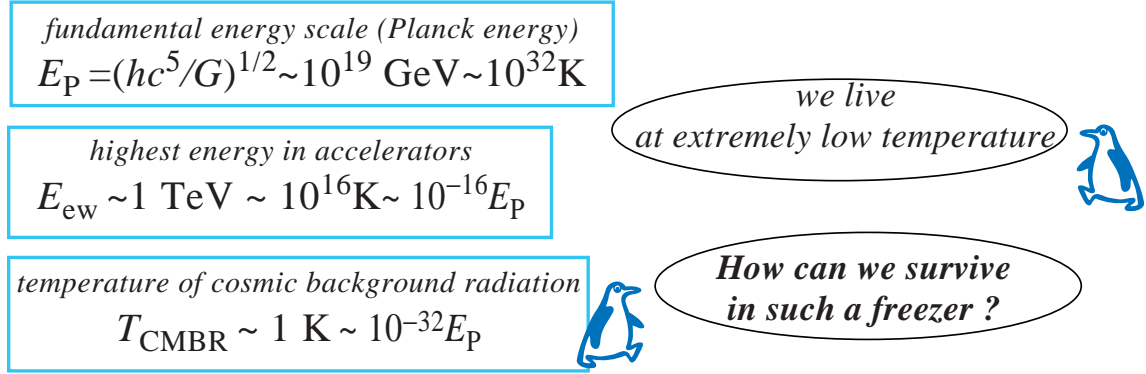}}
\caption{Characteristic energy scale in the vacuum of the ``natural Universe''
 is the Planck energy $E_P$. Compared to that energy, the high-energy physics and cosmology operate at extremely ultra-low temperatures.}  
\label{MainProblem} 
\end{figure}

That we survive in our Universe is not the result of the anthropic principle (the latter chooses the Universes which are fine-tuned for life but have an extremely low probability). On the contrary, this simply indicates that our Universe is also natural, and its vacuum is generic though it belongs to a universality class which is different from the Universes with massive particles. Indeed, the momentum space topology suggests that, both in relativistic quantum field theories and in fermionic condensed matter, there are several universality classes of quantum vacua (ground states) \cite{FrogNielBook,Book,Horava}. One of them contains vacua with trivial topology, whose fermionic excitations are massive (gapped) fermions. The natural mass  of these fermions is on the order of $E_{\rm P}$. 

The other classes contain gapless vacua. Their fermionic excitations live either near Fermi surface (as in metals), or near a Fermi point (as in superfluid $^3$He-A), or near some other topologically stable manifold of zeroes in the energy spectrum. The gaplessness of these fermions is protected by topology, and thus is not sensitive to the details of the microscopic (trans-Planckian) physics. Irrespective of the deformation of the parameters of the microscopic theory, the natural value of the gap in the energy spectrum of these fermions remains strictly zero.

\section{Emergent gravity in vacua with Fermi points}

\begin{figure}
\centerline{\includegraphics[width=1.0\linewidth]{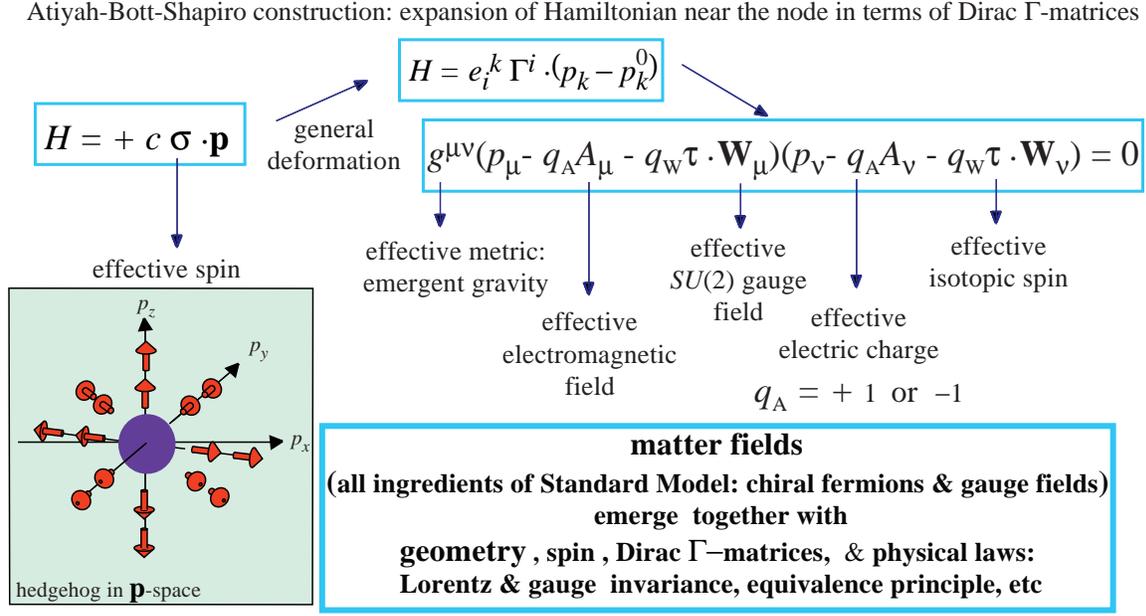}}
\caption{Relativistic quantum fields and gravity emerging near a Fermi point -- topologically protected hedgehog in momentum space. Spin of a right-handed fermion is directed along spines of the hedgehog}  
\label{EmergentPhysics} 
\end{figure}

For our Universe, which obeys Lorentz invariance,  only those vacua  are important that are either Lorentz invariant, or acquire Lorentz invariance as an effective  symmetry emerging at low energy. This excludes the vacua with Fermi surface and leaves the class of vacua with a Fermi point of chiral type (the hedgehog in momentum space, see Fig.~\ref{EmergentPhysics}), in which fermionic excitations behave as left-handed or right-handed Weyl fermions   \cite{FrogNielBook,Book}, and the class of vacua with the nodal point obeying $Z_2$ topology, where fermionic excitations behave as massless Majorana neutrinos   \cite{Horava}.

The advantage of the vacua with Fermi points is that practically all the main physical laws (except for quantum mechanics) can be considered as effective laws, which naturally emerge at low energy. This is the consequence of the  so-called Atiyah-Bott-Shapiro construction (see Ref.  \cite{Horava}),  which leads to the following general form of expansion for the Hamiltonian of fermionic quasiparticles near the Fermi point:
\begin{equation}
H=e_i^k\Gamma^i(p_k-p_k^0)+~{\rm higher~order~terms}~.
\label{Atiyah-Bott-Shapiro}
\end{equation}
Here the $\Gamma^i$ are Dirac matrices; the expansion parameters (the vector 
$p_k^0$ indicating the position of the Fermi point in momentum space and  the matrix $e_i^k$) depend on the space and time coordinates and thus are dynamic fields. This expansion demonstrates that close to the Fermi point, the  low-energy electrons  behave as relativistic Weyl fermions. 
The vector field  $p_k^0$ plays the role of   the effective $U(1)$ gauge field acting on these fermions. 
If $p_k^0$ is  the matrix field, it gives rise to effective non-Abelian gauge fields.
 The matrix field $e_i^k$ acts on the quasiparticles as a vierbein field, and thus describes dynamical gravity. As a result, close to the Fermi point,  matter fields 
(all ingredients of Standard Model: chiral fermions and Abelian and non-Abelian gauge fields) 
emerge  together with geometry, relativistic spin, Dirac  matrices,  and physical laws:  Lorentz and gauge  invariance, equivalence principle, etc.

 The existence of the Fermi point in the vacuum of our Universe is an experimental fact, since all our elementary particles, quarks and leptons, are chiral Weyl fermions. It is still not excluded that some of neutrinos are Majorana fermions, but this only changes the topological characteristic of the Fermi point. Does that mean gravity in our Universe is not fundamental? At the moment  there is no experimental evidence that  the fundamental theory must be abandoned. Moreover, the  existence of a Fermi point can be a property  of the fundamental theory too, in this case the second order and all higher order terms in Eq.(\ref{Atiyah-Bott-Shapiro}) are absent due to Lorentz symmetry. 

What about emergent gravity?  Can gravity be an effective low-energy phenomenon? 
Usually the induced Sakharov type gravity is abandoned using the argument that the induced cosmological constant is proportional to $E_{\rm P}^4$. From the renormalization group  point of view, this is the dominant term in the gravitational action, and its effect is that there is
no distance scale in the universe longer than $E_{\rm P}^{-1}$. So one has no
classical regime of gravity at all, and only a Planck scale Universe. However, in emergent scenarios this argument against effective gravity does not work: it is clear from the thermodynamic arguments that for any effective theory of gravity the natural value of $\Lambda$ is zero. This result does not depend on the microscopic structure of the vacuum  from which gravity emerges, and is actually the final result of the renormalization dictated by macroscopic physics (more on that see in Re. \cite{New}).

\section{Emergent gravity vs fundamental gravity}

\begin{figure}
\centerline{\includegraphics[width=1.0\linewidth]{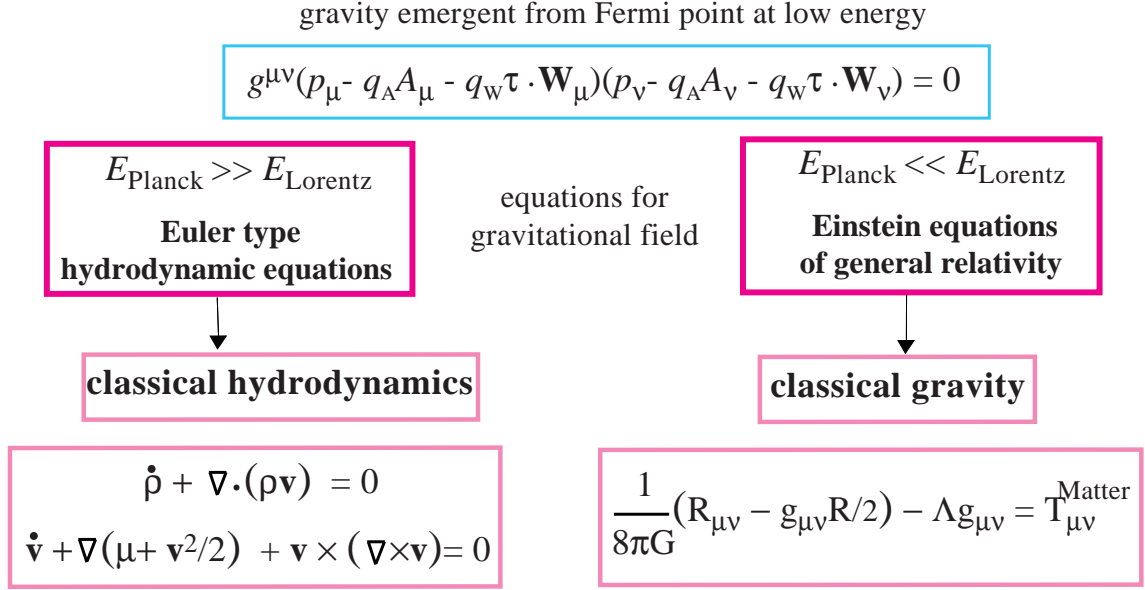}}
\caption{Equations for the metric field  $g_{\mu\nu}$ emerging near the Fermi point depend on hierarchy of ultraviolet cut-off's:
Planck energy scale $E_{\rm P}$ vs Lorentz violating scale $E_{\rm Lorentz }$.  }
\label{GravityAndHierarchy} 
\end{figure}

The Fermi-point scenario gives a particular mechanism of emergent gravity. This mechanism  has many consequences, and some of them can be used to falsify this scenario. So, let us assume that gravity is the low-energy effective theory emerging from the Atiyah-Bott-Shapiro construction. What are the consequences?

If gravity emerges from the Fermi point scenario, then:

(1) Gravity emerges together with matter (Fig. \ref{EmergentPhysics}).   This means that the so-called  ``quantum gravity'' must be the unified theory of the underlying quantum vacuum, where the gravitational degrees of freedom cannot be separated from all other microscopic  degrees of freedom, which give rise to the matter fields (fermions and gauge fields).

(2) Fermionic matter, which emerges together with gravity, consists of Weyl fermions. This agrees with the fermionic content of our Universe, where the elementary particles are left-handed and right-handed quarks and leptons.
 
(3) Gravity cannot be quantized. Gravity is the result of an up-down procedure: it is the low-energy macroscopic classical output of the high energy microscopic quantum vacuum. The inverse down-up procedure from classical to quantum gravity is highly restricted.  The first steps in quantization are allowed: it is possible to quantize gravitational waves to obtain their quanta -- gravitons; it is possible to obtain some quantum corrections to Einstein equation; to extend classical gravity to the semiclassical and stochastic \cite{Hu} levels,  etc.  But one cannot  cannot  obtain ``quantum gravity'' by full quantization of Einstein equations.

(4) Effective gravity may essentially differ from fundamental
gravity even in principle. Since in effective gravity general covariance is lost at high energy, metrics which for the low-energy observers appear equivalent, since they can be transformed into each other by mathematical coordinate transformation, need not be equivalent physically. As a result, in emergent gravity some metrics, which are natural in general relativity, are simply forbidden. 
For example, emergent gravity is not able to incorporate the geodesically-complete Einstein Universe with spatial section $S^3$  \cite{KlinkhamerVolovikCoexisting}.  It, therefore, appears that the original static $S^3$ Einstein Universe \cite{Einstein} can exist only within the context of fundamental general relativity. Some coordinate transformations in GR are not allowed in emergent gravity: these are either singular transformations of the original  coordinates, or transformations which remove some parts of spacetime (or add the extra parts).
The non-equivalence of different metrics is especially important
in the presence of an event horizon. 
For example, in emergent gravity the Painlev\'e-Gullstrand metric is more appropriate for the description of a black hole, than the Schwarzschild metric, which is (coordinate) singular at the horizon.

(5) The Universe is naturally flat. In fundamental general relativity, the isotropic and homogeneous Universe means the space with constant curvature. In emergent gravity with an effective metric, the isotropic and homogeneous Universe corresponds to flat space. In general relativity the flatness of the  Universe requires either fine tuning or  inflationary scenario in which the curvature term is exponentially suppressed if the exponential inflation of the Universe irons out curved space to make it extraordinarily flat. The observed flatness of our Universe is in favor of emergent gravity. 

\begin{figure}
\centerline{\includegraphics[width=1.0\linewidth]{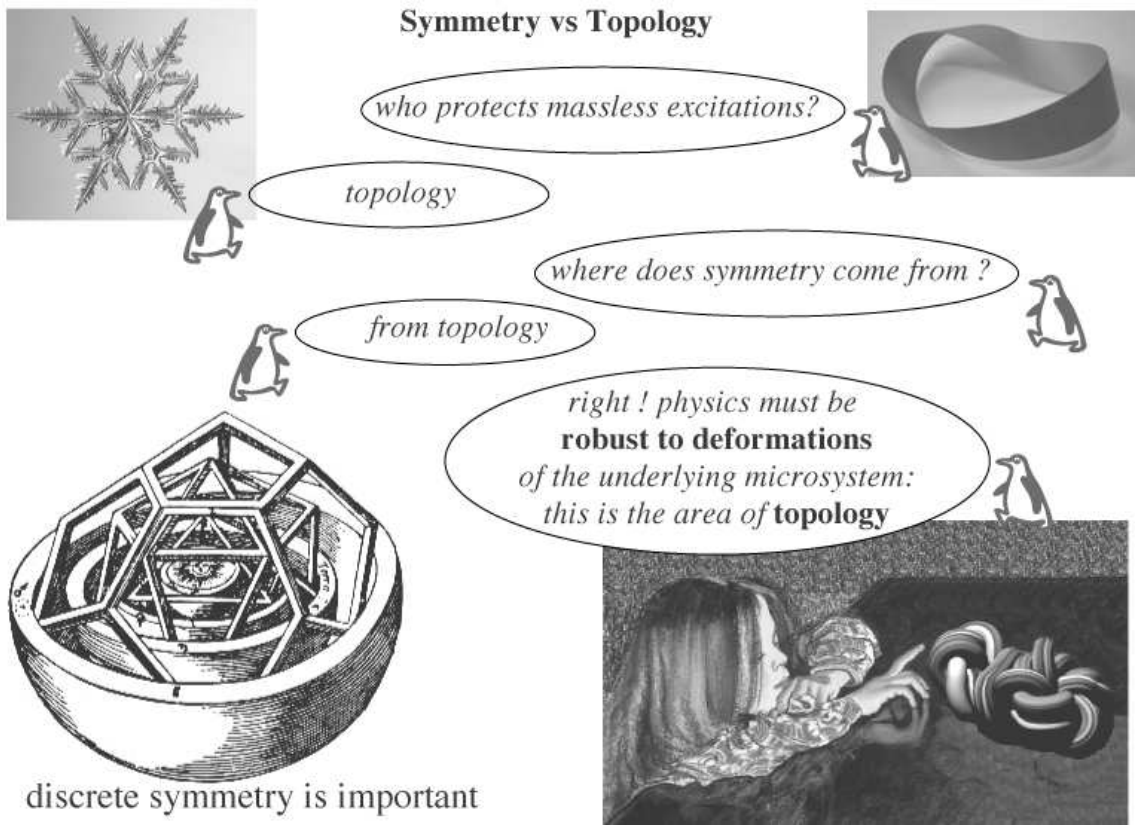}}
\caption{Momentum-space topology is the main source of massless elementary particles. But it must be accompanied by discrete symmetries between Fermi points (see Fig. \protect\ref{TwoScenarios}). {\it bottom right}: from ``Knots in art''  by  Piotr Pieranski. }  
\label{SymVsTopology.eps}
\end{figure}

(6) The cosmological constant is naturally small or zero. In general relativity, the cosmological constant  is an arbitrary constant, and thus its smallness requires fine-tuning. Thus observations are in favor of emergent gravity. The unsolved problem is: what is the physical mechanism of relaxation of $\Lambda$ towards zero? (See Ref. \cite{Barcelo} and references therein.) The present small value of $\Lambda$ indicates that the Universe is so close to  equilibrium that the current relaxation rate is very slow.

(7) In the Fermi point scenario space-time is naturally  4-dimensional.  This is a fundamental property of the Fermi-point topology, which as distinct from the string theory does not require the higher-dimensional space-times. 

\begin{figure}
\centerline{\includegraphics[width=0.7\linewidth]{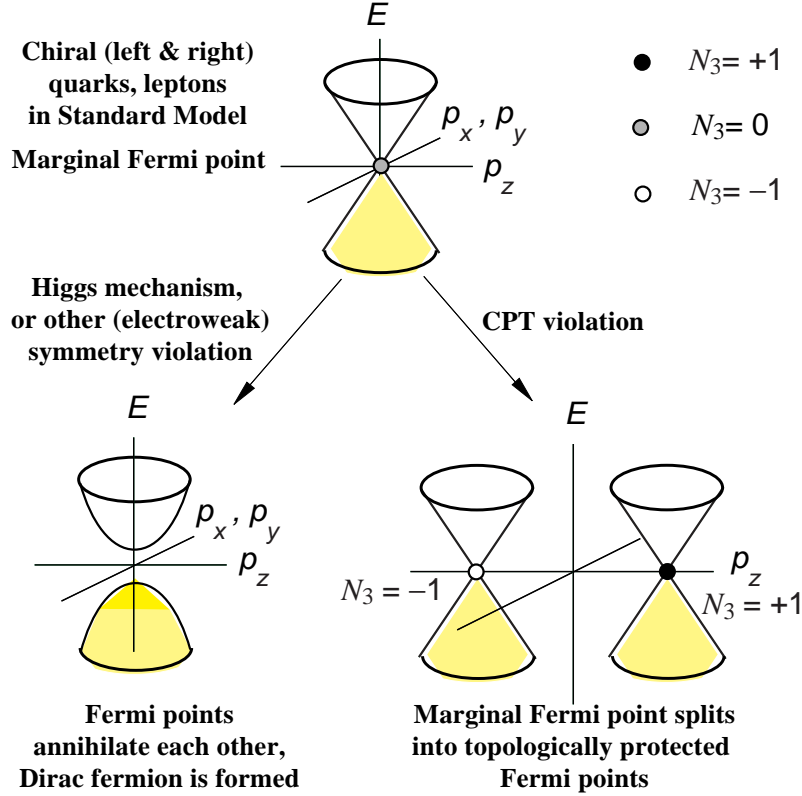}}
\caption{({\it top}) In Standard Model the Fermi points with positive $N_3=+1$ and negative $N_3=-1$ topological charges are at the same point ${\bf p}=0$. It is the discrete symmetry between the Fermi points which prevents their mutual annihilation. When this symmetry is violated or spontaneously broken, there are two topologically different scenarios: ({\it bottom left}) either Fermi point annihilate each other and  Dirac mass is formed;  ({\it bottom right})  or Fermi points split
\protect\cite{Splitting}. It is possible that actually the splitting exists at the microscopic level,  but in our low energy corner we cannot observe it because of the emergent gauge symmetry: in some cases  splitting can be removed by  gauge transformation.}  
\label{TwoScenarios} 
\end{figure}

(8) The underlying physics must contain discrete symmetries (Fig. \ref{SymVsTopology.eps}). Their role is extremely important. The main role is to prohibit the cancellation of Fermi points with opposite topological charges (see Fig. \ref{TwoScenarios}). As a side effect, in the low-energy corner discrete symmetries are transformed into gauge symmetries, and give rise to gauge fields. They also reduce the number of  massless gauge bosons. To justify the Fermi point scenario, one should find the discrete symmetry which leads in the low energy corner to one of the GUT or Pati-Salam models.

\begin{figure}
\centerline{\includegraphics[width=1.0\linewidth]{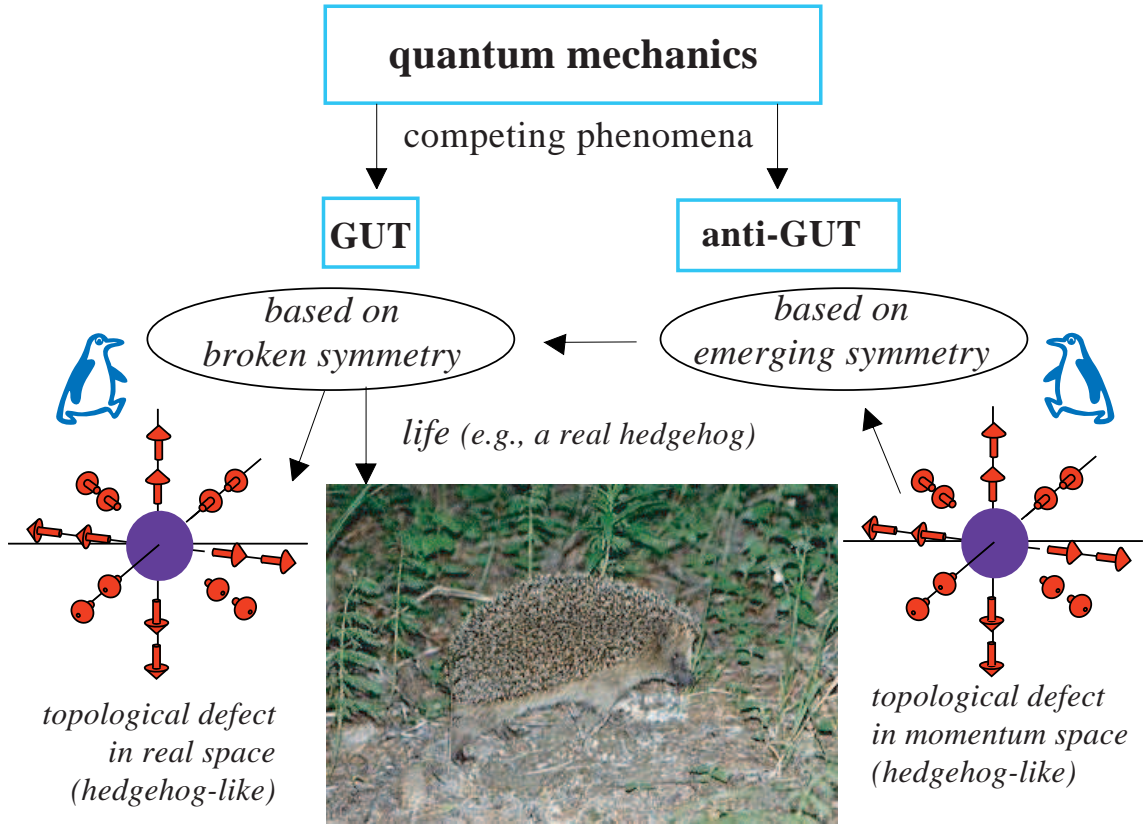}}
\caption{From history of hedgehogs, or three elements of modern physics: (i) quantum mechanics (or quantum field theory); (ii) Grand Unification based on the phenomenon of broken symmetry at low energy (GUT symmetry is restored when the Planck energy scale is approached from below); and (iii) anti-GUT based on the opposite phenomenon -- GUT symmetry gradually emerges when the Planck energy scale is approached from above.  A hedgehog-like  topological defect  in momentum space -- the Fermi point -- gives rise to symmetry emergent  at Planck-GUT scales. In turn,  symmetry breaking  occurring at lower energy,  gives rise to topological defects in real space (e.g., a hedgehog-like object)  and life (e.g., a real hedgehog).}  
\label{TriKita} 
\end{figure}

(9) Lorentz symmetry must persist well above the Planck energy. The requirement for two well separated energy scales follows from the high precision of physical laws in our Universe \cite{Bjorken2001}, and it represents the most crucial test of the emergent scenario.  In the case when the Lorentz violating scale $E_{\rm Lorentz }< E_{\rm P}$, the metric field does not obey Einstein equations; instead it is governed by the hydrodynamic type equations (see Fig.~\ref{GravityAndHierarchy}). The Einstein equations emerge in the limit $E_{\rm Lorentz }\gg E_{\rm P}$, and their accuracy  is determined by the small parameter $E_{\rm P}^2/E_{\rm Lorentz }^2\ll 1$.  For example, in the Frolov-Fursaev version of Sakharov induced gravity \cite{FrolovFursaev1998}, the ultraviolet cut-off is much larger than the Planck energy, and Einstein equations are reproduced. The observed bounds on the violation of Lorentz symmetry can be  obtained from ultra-high-energy cosmic rays. For example,  according to conservative estimations the relative value of  the Lorentz violating terms in the Maxwell equations is below $10^{-18}$ \cite{KlinkhamerRisse}. This suggests that  $E_{\rm Lorentz } > 10^9E_{\rm P}$, which is in favor of emergent scenario.

 All this implies that physics continues far beyond the Planck scale, and this opens new  possibilities for construction of microscopic theories. Since in the Fermi point scenario bosons are composite objects, the ultraviolet  cut-off is different for fermions and bosons \cite{KlinkhamerVolovikMerging} (a similar situation occurs in condensed matter \cite{Chubukov}). The smaller (composite) scale can be associated with $E_{\rm P}$, while
the "atomic" structure of the quantum vacuum will be only revealed at
much higher Lorentz-violating scale $E_{\rm Lorentz}$. The opposite situation, when fermionic degrees of freedom emerge in the underlying bosonic quantum vacuum, is possible (see \cite{Kitaev,Wen} and especially Ref.  \cite{YueYu}, where the Fermi points emerge in some model of spins  on a three-dimensional lattice), but at the moment the generic mechanism for the emergence of gravity and matter in bosonic vacuum has not yet been found.

 (10) Finally, what about quantum mechanics? Actually both schemes for the classification of
quantum vacua: by symmetry (GUT scheme) and  by topology  in momentum space (anti-GUT scheme)
are based on quantum mechanics (Fig. \ref {TriKita}). While general relativity is assumed to be as fundamental as quantum mechanics, emergent gravity with its emergent metric of the effective low-energy space-time is a secondary phenomenon. It is the byproduct of quantum field theory or of many-body quantum mechanics. As a result, in the Fermi-point scenario there are no principle contradictions  between quantum mechanics and gravity. Due to the same reason, emergent gravity cannot be responsible for the issues related to foundations of quantum mechanics, and in particular for the collapse of the wave function. Also,  item (9) implies that if  quantum mechanics is not fundamental, the scale at which it emerges is far beyond the Planck scale.

I thank Frans Klinkhamer for fruitful discussions. This work has been supported in part by the European Science Foundation network
programme "Quantum Geometry and Quantum Gravity" and by the Russian
Foundation for Fundamental Research.

\end{document}